\documentclass[]{spie}  

\usepackage{amsmath,amsfonts,amssymb}
\usepackage{graphicx}
\usepackage[colorlinks=true, allcolors=blue]{hyperref}
\usepackage{rotating}
\usepackage{cite}

\title{The LiteBIRD mission to explore cosmic inflation}

\include{authors_affiliations_LB_2024.05_general_spie}

\authorinfo{Further author information: (Send correspondence to T.G.) E-mail: tommaso.ghigna@kek.jp}

\begin{document} 
\maketitle

\begin{abstract}
\textit{LiteBIRD}, the next-generation cosmic microwave background (CMB) experiment, aims for a launch in Japan's fiscal year 2032, marking a major advancement in the exploration of primordial cosmology and fundamental physics. Orbiting the Sun-Earth Lagrangian point L2, this JAXA-led strategic L-class mission will conduct a comprehensive mapping of the CMB polarization across the entire sky. During its 3-year mission, \textit{LiteBIRD} will employ three telescopes within 15 unique frequency bands (ranging from 34 through 448 GHz), targeting a sensitivity of 2.2\,$\mu$K-arcmin and a resolution of 0.5$^\circ$ at 100\,GHz. Its primary goal is to measure the tensor-to-scalar ratio $r$ with an uncertainty $\delta r = 0.001$, including systematic errors and margin. If $r \geq 0.01$, \textit{LiteBIRD} expects to achieve a $>5\sigma$ detection in the $\ell=2$--10 and $\ell=11$--200 ranges separately, providing crucial insight into the early Universe.
We describe \textit{LiteBIRD}'s scientific objectives, the application of systems engineering to mission requirements, the anticipated scientific impact, and the operations and scanning strategies vital to minimizing systematic effects. We will also highlight \textit{LiteBIRD}'s synergies with concurrent CMB projects.
\end{abstract}

\keywords{\textit{LiteBIRD}, cosmic inflation, cosmic microwave background, $B$-mode polarization, primordial gravitational waves, quantum gravity, space telescope}

\section{SCIENCE AND MISSION OVERVIEW}
\label{sec:intro}
\textit{LiteBIRD}\cite{Hazumi2008AIP, Hazumi2011PTPS, Hazumi2012SPIE, Matsumura2014JLTP, Matsumura2014SPIE, Suzuki2018JLTP, Hazumi2020SPIE, Hazumi2021PTEP}, the Lite (Light) satellite for the study of $B$-mode polarization and Inflation from cosmic background Radiation Detection, is a space mission aimed at exploring primordial cosmology and fundamental physics through the study of $B$-mode polarization and inflation (a phase of rapid expansion in the early Universe) in the cosmic microwave background (CMB)\cite{kamionkowski97, zaldarriaga_seljak97}. The project began with preliminary conceptual studies in 2008 and was proposed as a candidate for a JAXA large-class (L-class) mission in 2015. The launch plan is to make use of the newly developed H3 rocket that recently entered service during a second launch attempt on 17 February 2024 from Tanegashima Space Center in Western Japan.

\begin{figure}[tp]
    \centering
    \includegraphics[width=1\textwidth]{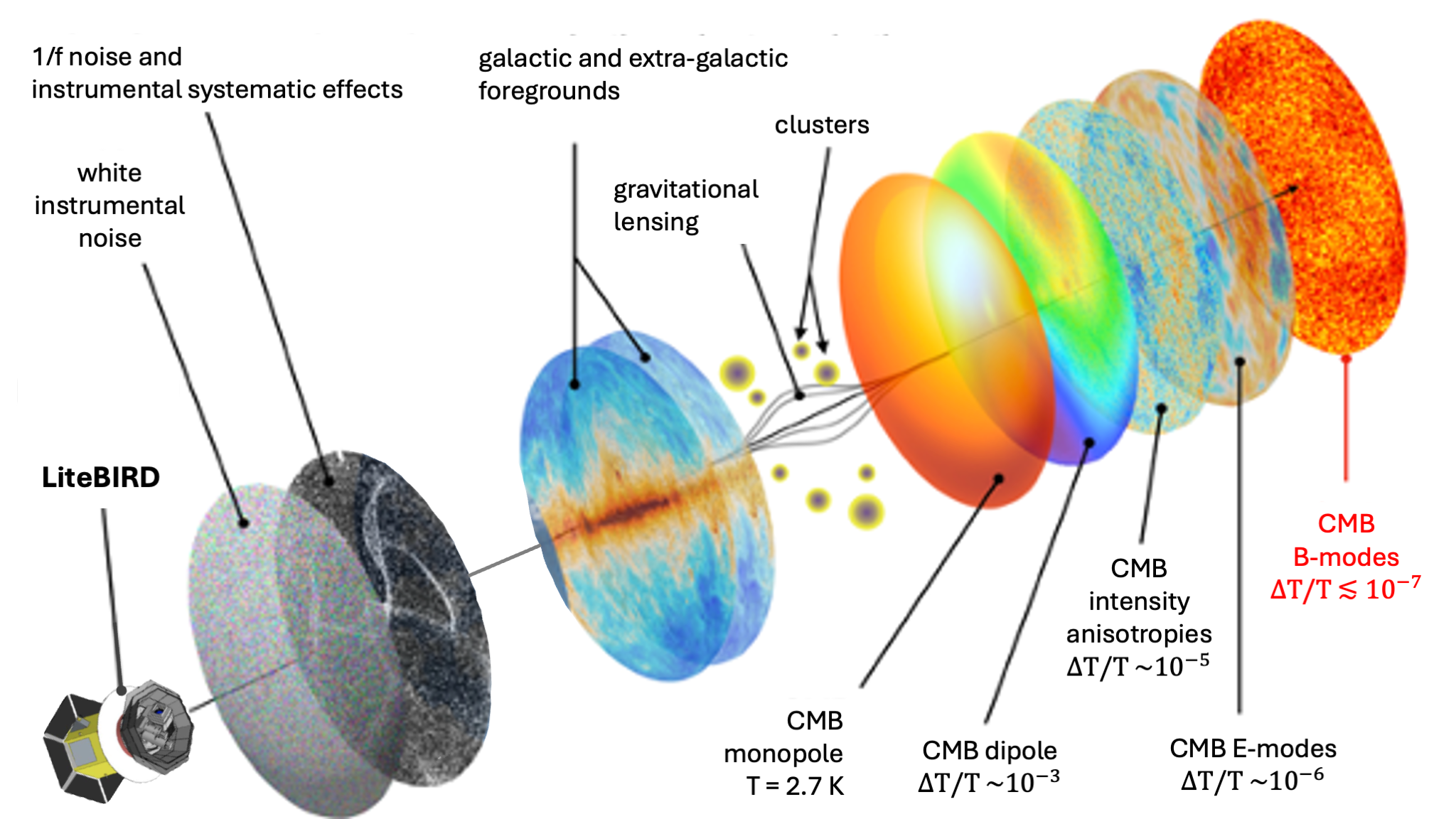}
    \caption{Illustration of the additive components contributing to the \textit{LiteBIRD} data. These include instrumental noise, systematic effects, galactic and extragalactic sources, in addition to the primary scientific target: the faint primordial $B$-mode signal. Credits: Josquin Errard and \citenum{errard_erc}.}%
    \label{fig:litebird}
\end{figure}

\textit{LiteBIRD} successfully completed the mission definition review (MDR) in early 2024, after which the project will enter JAXA's phase A (pre-project phase) after undergoing the project readiness review (PRR). This phase is expected to last 3 years until the system definition review (SDR). The mission boasts a collaborative effort with over 250 researchers from Japan, North America, and Europe, including experts in CMB experiments, X-ray satellite missions, high-energy physics and astronomy. This large collaborative effort and broad expertise is needed because of the complexity of the challenge of measuring CMB $B$ modes, as highlighted by the illustration in Figure~\ref{fig:litebird}.

As stated by the experiment top-level scientific requirement, \textit{LiteBIRD}'s main target is the detection of CMB $B$ modes, with the goal of constraining cosmic inflation with a total uncertainty ($\delta r = 0.001$), which will allow us to provide definite statements on many inflationary models. At the same time, \textit{LiteBIRD}, by providing high-sensitivity full-sky data across a wide frequency range, has the potential to shed new light on some hints of possible physics beyond the standard model that have been reported by recent CMB experiments.

\textit{LiteBIRD} is not alone in its quest for CMB $B$ modes. In the next decade several ground-based experiments will observe the sky with the same goal\cite{simonsObs2019,cmbs4}. CMB $B$ modes are regarded as the most sensitive tracer of cosmological inflation, which remains one of the big unanswered questions about the origin of all structure. Ground and space experiments should not be seen as competing with each other in this quest, but rather as complementary. The strength of a space mission resides in the availability of the full sky on short timescales, where large angular scales can be accessed efficiently, while ground-based experiments can target the smaller scales that are necessary to delens the data in order to achieve higher sensitivity to the primordial $B$-mode signal. For these reasons, the \textit{LiteBIRD} and CMB-S4 \cite{cmbs4} collaborations have recently established a common working group to test possible ways of combining the strengths of the two experiments with the goal of boosting the overall scientific output.  

The SPIE 2024 conference includes detailed discussions on many aspects of \textit{LiteBIRD}\cite{Kaneko2024SPIE, Roudil2024SPIE, deHaan2024SPIE, Raum2024SPIE, Farias2024SPIE, Ghigna2024SPIE, Takaku2024SPIE, Akizawa2024SPIE, Maestre2024SPIE, Matsuda2024SPIE, Micheli2024SPIE, Columbro2024SPIE, Takakura2024SPIE, Matsumura2024SPIE, Occhiuzzi2024SPIE, Takahashi2024SPIE, Mot2024SPIE, Miura2024SPIE, Oguri2024SPIE, Maffei2024SPIE, Savini2024SPIE, Stever2024SPIE, Chahadih2024SPIE, Hastanin2024SPIE, deHaan2024SPIEgeneral}. From spacecraft components and operation to the physics of superconductive TES bolometers, from optical modeling and measurements to discussion of systematic effects and mitigation techniques, with this paper providing a concise overview of the mission and an update about schedule and some of the mission parameters after the end of MDR. This overview serves as an introduction to other \textit{LiteBIRD} contributions and will focus on mission requirements, global design and operation, and scientific outcomes. 
In Section~\ref{sec:requirement} we recap the top-level scientific requirements and explain the requirement flow-down philosophy. In Section~\ref{sec:schedule} we give an update of the mission schedule following the recent outcome of the MDR. In Sections~\ref{sec:spacecraft} and \ref{sec:operation} we give an overview of the spacecraft and the payload and mission operations, respectively. In Section~\ref{sec:science} we present the expected scientific outcomes of the mission, and in Section~\ref{sec:conclusions} we give a summary and conclusions.

\section{REQUIREMENTS FLOW-DOWN}
\label{sec:requirement}
\textit{LiteBIRD}'s primary science target is the large angular scale CMB $B$ modes. This polarized signal can be used as a tracer of the energy of inflation. The standard parameter is the tensor-to-scalar ratio ($r$) that measures the amplitude of tensor modes generated during inflation relative to the amplitude of scalar modes. At present only upper limits on $r$ have been set, with the tightest constraint being $r<0.032$ from a combined analysis of the Planck \cite{Planck2018I} and BICEP \cite{Bicep2018limit} data sets\cite{Tristram2022}. This upper limit is already approaching the level where the primordial signal is expected to be fully dominated by the lensing signal
(due to conversion of $E$ modes to $B$ modes through gravitational lensing from large-scale structures along the line of sight) for angular scales $\ell\gtrsim100$. Therefore, only an experiment sensitive to larger scales $\ell\lesssim100$ (which is very challenging from the ground) can hope to detect the primordial signal without requiring a delensing analysis. The goal in the next decade, not just for \textit{LiteBIRD}, is to lower the limit by more than one order of magnitude to $r\sim0.001$. This is not a purely arbitrary choice, but is physically motivated. At this level of uncertainty on $r$ we should be able to confirm or rule out single-field models for which the characteristic scale of field-variation of the inflationary potential is larger than the reduced Planck mass\cite{Linde_2017}.
\begin{figure}[tp]
    \centering
    \includegraphics[width=1\textwidth]{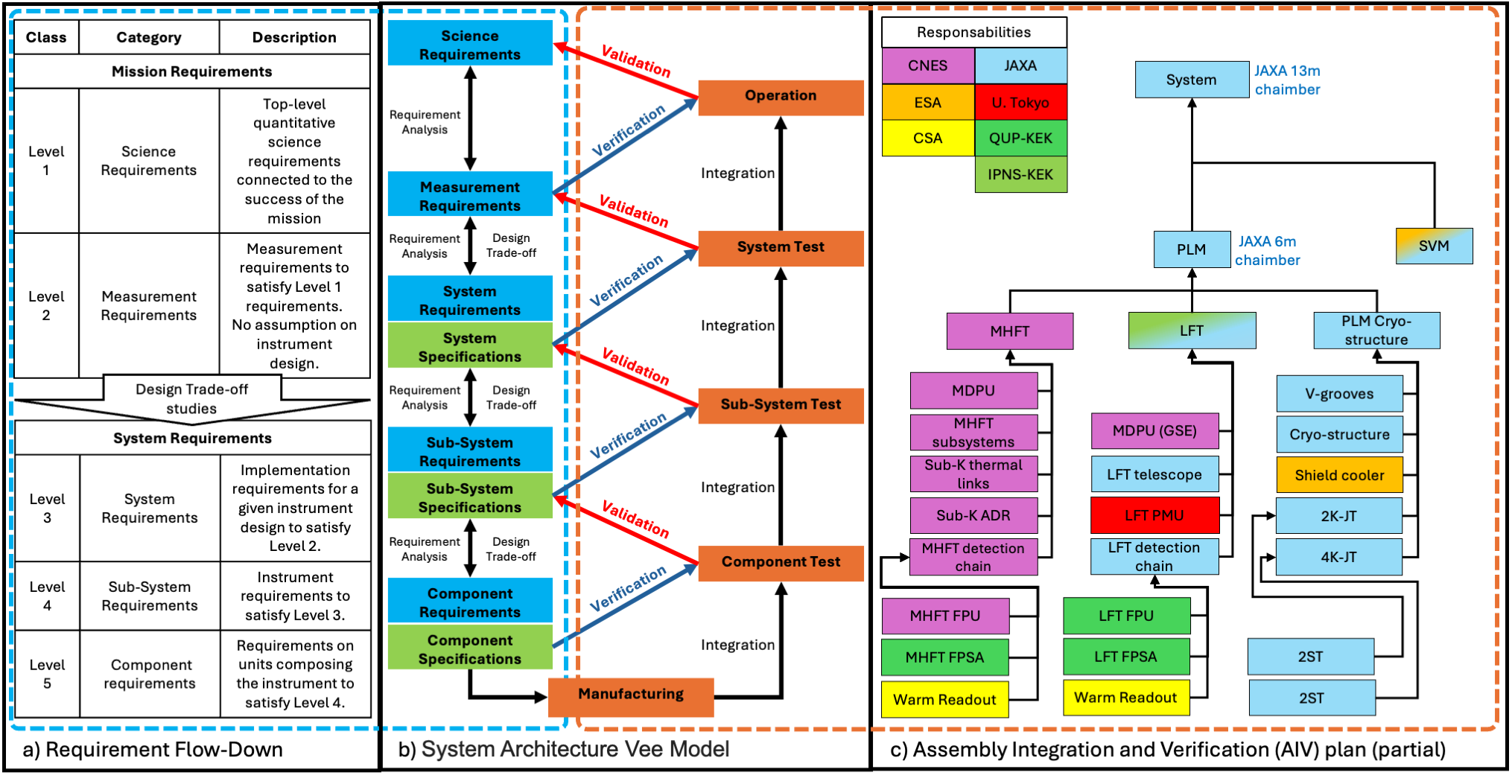}
    \caption{Simplified schematic of the requirements flow-down (a) and verification plan (c), based on systems engineering concepts (b). We highlight the connection between requirements and the validation and verification cycle at each stage of the system development. For more details on this topic, see \citenum{Kaneko2024SPIE}.}%
    \label{fig:req_aiv}
\end{figure}

In order to achieve the aforementioned goal, the \textit{LiteBIRD} collaboration has decided to set two top-level scientific requirements (Level 1) to achieve the goal of the mission \cite{Hazumi2021PTEP, Hazumi2020SPIE}. 
\begin{enumerate}
    \item \textbf{Total measurement uncertainty for tensor-to-scalar ratio r:} The mission shall measure $r$ with a total uncertainty of $\delta r < 0.001$. This value shall include contributions from instrument statistical noise fluctuations, instrumental systematics, residual foregrounds, lensing $B$ modes, and observer bias, and shall not rely on future external data sets.
    \item \textbf{Detection sensitivity for non-zero tensor-to-scalar ratio r:} The mission shall obtain full-sky CMB linear polarization maps for achieving $>5\sigma$ significance using $2\leq\ell\leq10$ and $11\leq\ell\leq200$ separately, assuming $r=0.01$ and adopting a fiducial optical depth of $\tau=0.05$ for the calculation.
\end{enumerate}

From these two top-level scientific requirements we follow a requirement analysis or flow-down to impose requirements on each component of the mission that allow us to satisfy the scientific requirements. While the lower level requirements are still being actively defined by the collaboration, the philosophy and framework are well established, as shown in Figure~\ref{fig:req_aiv}. This is a simplified schematic of the requirement flow-down and verification plan based on systems engineering concepts\cite{incose}. Starting from the left side of Figure~\ref{fig:req_aiv}a, we show the requirement flow-down philosophy for \textit{LiteBIRD}. Level~1 describes the scientific requirements of the mission as mentioned above. Level~2 imposes measurement requirements to satisfy the scientific requirement at Level~1 without any assumptions on the instrument design. We have identified 11 Level~2 measurement requirements,
which we will not report here because an extensive description can be found in \citenum{Hazumi2020SPIE, Hazumi2021PTEP}. After the Level~2 requirements, we have to start making some decisions on the instrument design to define lower level requirements (system, sub-systems and components). Many trade-off studies on the design of the \textit{LiteBIRD} system and sub-systems have been presented in the past years with the goal of optimizing the instrument performance without violating the requirements from the level above.

\begin{figure}[tp]
    \centering
    \includegraphics[width=1\textwidth]{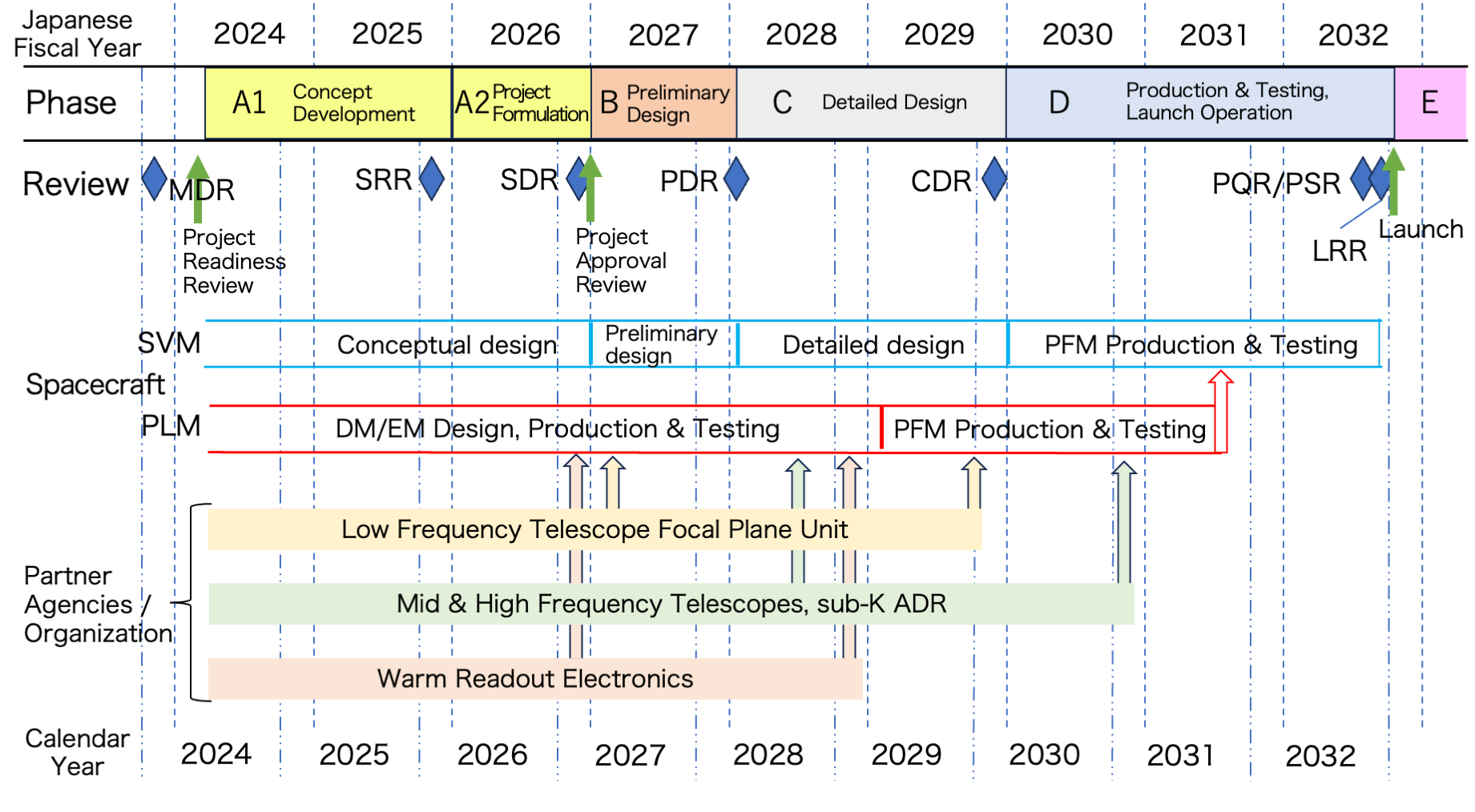}
    \caption{\textit{LiteBIRD} pre-flight schedule as of 2024. The top rows show the correspondence between the project phases and Japanese Fiscal Years (JFY). At the bottom the corresponding calendar year is indicated. Here we show the cadence of the major JAXA reviews, the main milestones and the stage of development of the service module (SVM) and payload modules (PLM), highlighting in particular the model-based development approach. Note, for example, that the PLM will be developed in three stages: design model (DM); engineering model (EM); and PFM (proto-flight model). This approach is adopted in an effort to reduce risks.}%
    \label{fig:schedule}
\end{figure}

Figure~\ref{fig:req_aiv}b shows the relationship between requirement flow-down and system assembly, integration and verification (AIV). Figure~\ref{fig:req_aiv}c shows a simplified AIV scheme color-coded to highlight the responsible agency or institute.

\section{MISSION SCHEDULE}
\label{sec:schedule}
As mentioned in the introduction, the \textit{LiteBIRD} project originated more than 15 years ago, with initial conceptual studies in 2008, and was officially proposed as a candidate for JAXA large-class mission in 2015. In 2018 \textit{LiteBIRD} completed a two-year long pre-phase A2 concept development phase that culminated in its selection in 2019 as the second JAXA strategic large-class mission\cite{Hazumi2020SPIE}.

Since 2019 the project has been in JAXA pre-phase~A2 stage to continue conceptual studies, which lasted until early 2024 and ended after a positive conclusion of the mission definition review (MDR). As of today, we are expecting completion of the project readiness review (PRR) and the official start of the JAXA pre-project phase (Phase~A). As illustrated in Figure~\ref{fig:schedule}, the new launch date officially announced by JAXA, is expected to be at the end of Japanese Fiscal Year (JFY) 2032, which corresponds to the first quarter of calendar year 2033.

In Figure~\ref{fig:schedule} we highlight the cadence of the major JAXA reviews, starting from the recent MDR. The other main JAXA reviews are the system requirement review (SRR), the system definition review (SDR), which will mark the end of phase~A
and beginning of the project phase, the preliminary design review (PDR), which will mark the end of phase~B, the critical design review (CDR), the post-qualification test and pre-shipment reviews (PQR/PSR) that mark the end of phase D and the launch readiness review (LRR). We also show the main milestones and the stage of development of the service module (SVM) and payload modules (PLM), highlighting in particular the model-based development approach. For example the PLM (and all its components) will be developed in three main stages: design model (DM); engineering model (EM); and flight model (FM). 

\begin{figure}[tp]
    \centering
    \includegraphics[width=1\textwidth]{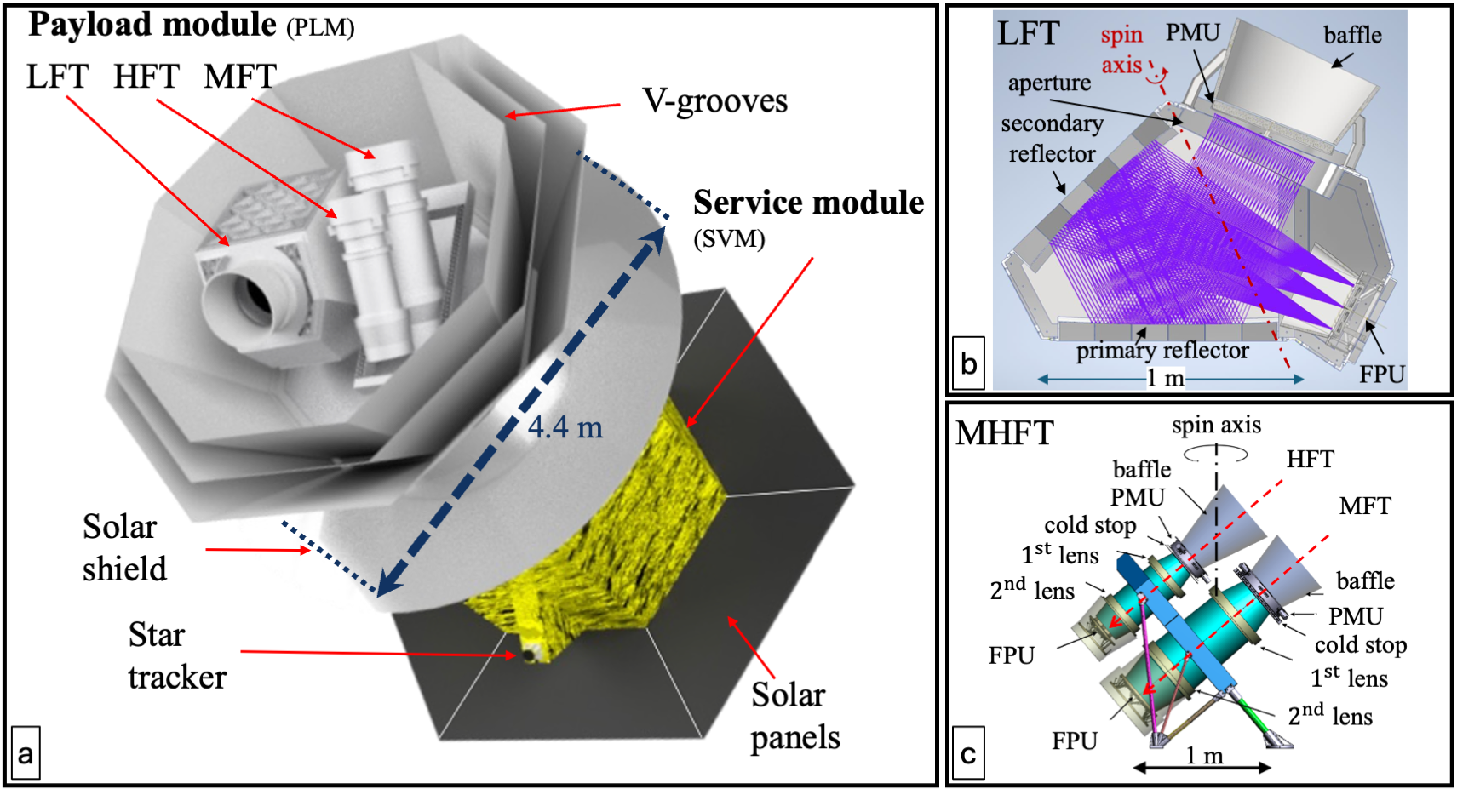}
    \caption{(a) Schematic of the spacecraft. We highlight the payload module (PLM) and the service module (SVM) with the main components. (b) The reflective low-frequency telescope (LFT) and (c) the refractive mid- and high-frequency telescopes (MHFT). Figures adapted from \citenum{Hazumi2021PTEP, Oguri2022SPIE, Montier2020SPIE}.}%
    \label{fig:spacecraft}
\end{figure}

This approach is adopted in an effort to reduce risk. To further minimise risk for certain critical (like the focal plane unit or FPU) or low TRL (technology readiness level) components, we also introduce an extra step corresponding in time with the STM stage, called the demonstration model (DM). 

\section{SPACECRAFT AND PAYLOAD}
\label{sec:spacecraft}
\textit{LiteBIRD}'s spacecraft structure features an axisymmetric design optimized for spinning. Placing the telescopes and solar panels at opposing ends, with the payload module (PLM) on top and the solar panels below, perpendicular to the spin axis. To mitigate interference with the telescopes, the high-gain antenna is positioned on the underside, opposite the mission instruments, pointing towards Earth. Figure~\ref{fig:spacecraft} provides a visual representation of the spacecraft configuration.

The decision to have the entire spacecraft spin, rather than employing a slip-ring to rotate the PLM only, stemmed from various considerations. These included managing significant heat dissipation in the PLM and reducing single-point failure risks. The mechanical coolers in the PLM dissipate a substantial amount of heat, necessitating a radiator size that fits in the service module (SVM). Additionally, transferring heat from the spinning PLM to a non-spinning SVM via a slip-ring was deemed challenging due to concerns about potential micro-vibrations and increased detector noise. Consequently, rotation of the entire spacecraft has been adopted as the baseline.

The spacecraft's center accommodates a truss structure, which transfers the PLM launch load to the rocket and houses the fuel tank for efficient use of the inner space. Electrical components of both the SVM and PLM are located within the SVM side panels, with PLM components preferably positioned on the upper parts and SVM 
components on the lower parts. Radiators on the outer sides of the upper panel parts dissipate heat from the PLM, including that generated by mechanical coolers and electronic boxes.
 
\begin{figure}[tp]
    \centering
    \includegraphics[width=1\textwidth]{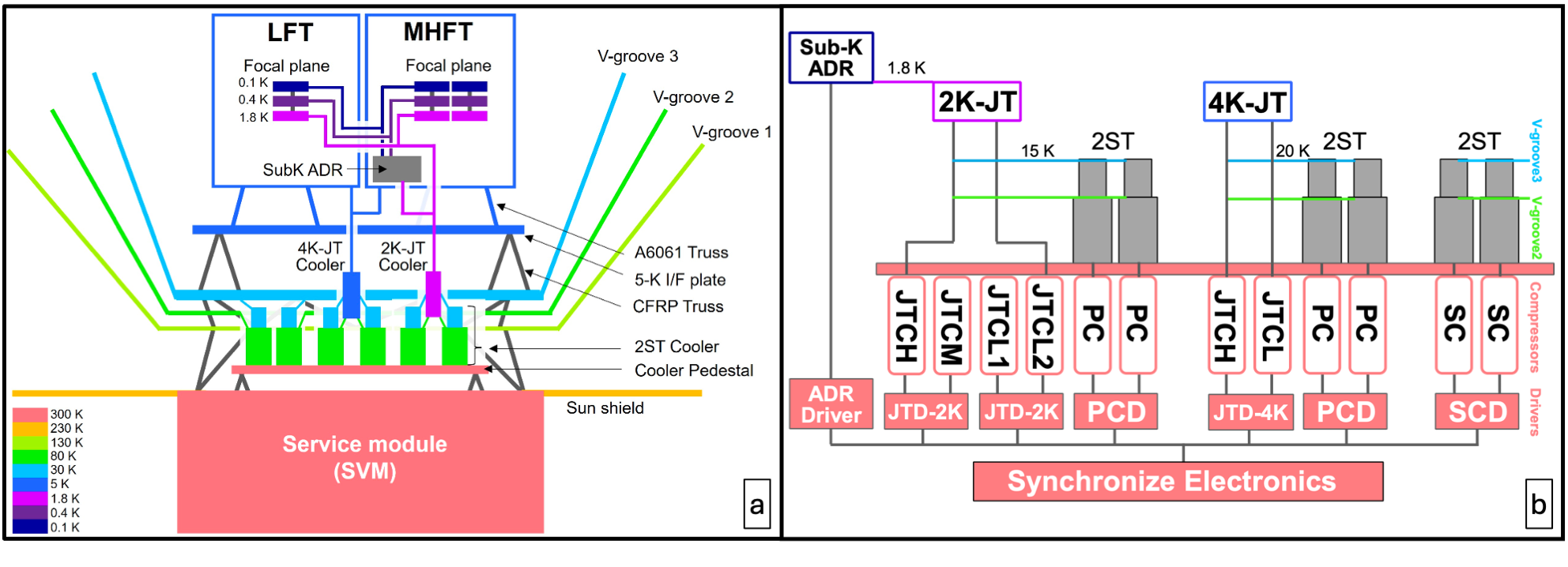}
    \caption{(a) Diagram depicting the thermal environment of the payload module (PLM). The Sun-shield delimits the service module (SVM) from the PLM and reduces the thermal loading on the PLM. A series of three V-grooves is employed in combination with two-stage Sterling coolers (2ST) to cool the PLM below 30\,K. Active cooling (combination of Joule-Thomson (JT) and adiabatic demagnetization refrigerators) is used to set the temperature of the coldest stages all the way to the 100\,mK required for the focal plane units (FPU). (b) Diagram of the cooling system. Figures are adapted from \citenum{Odagiri2023}. Most acronyms are explained througout the text, but here is a list of those that are not explained elsewhere: JTCH = JT compressor for high-pressure stage; JTCM = JT compressor for medium-pressure stage; JTCL = JT compressor for low-pressure stage; JTD = JT cooler driver; PC = precooler; PCD = precooler driver; SC = shield cooler; and SCD = shield cooler driver. }%
    \label{fig:plm}
\end{figure}

While spinning, \textit{LiteBIRD} employs an attitude control system to stabilize the satellite for precise attitude control and determination. The low spin rate (nominal 0.05\,rpm) allows for this operational mode. The spacecraft's estimated total weight, including fuel, is 3200\,kg, and requires an estimated 3.7\,kW to power the whole system. The data transfer occurs at a rate of 10\,Mbits\,s$^{-1}$ using the X-band antenna positioned on the opposite side of the spacecraft with respect to the payload and facing Earth. We estimate a total of 17.9\,GB of scientific and house-keeping data transmitted daily. Two JAXA ground stations are being currently considered for communication with the \textit{LiteBIRD} spacecraft, a planned successor station to Uchinoura Space Center 34 meters antenna and Misasa deep space station. The former may be used for telemetry and commands and down-link of the mission data, and the latter for telemetry and commands. Use of ESA (European Space Agency) stations for initial and critical operations is also being considered. 

All the parameters and information discussed here (summarized in Table~\ref{tab:paramters}) are up to date as of early 2024 (post MDR), although they may be subject to change, since the spececraft's conceptual design could still evolve and more detailed estimates will be available. 

Figure~\ref{fig:spacecraft} provides an overview of \textit{LiteBIRD}'s three telescopes mounted on the payload module. The \textit{LiteBIRD} payload module comprises three telescopes operating at low- (LFT), medium- (MFT), and high-frequencies (HFT), each with its focal planes and cryostructure cooled to 0.1\,K. The payload module also includes the global cooling chain shown in Figure~\ref{fig:plm}. This includes a Sun shield (which also marks the interface between the SVM and the PLM), the three V-groves and a series of 2STs that are necessary to cool the PLM from 300\,K to below about 30\,K. A combination of 2STs, Joule-Thomson coolers, and adiabatic demagnetization refrigerators is employed to cool the coldest stages to the sub-Kelvin temperatures required by the detector and readout components of the three focal-plane units. 

The reason for dividing the detectors between three different telescopes stems from technical challenges related to sensitivity, optical properties, stability, and compactness across a wide frequency range of 34 to 448\,GHz. One of the main challenges is robust control of systematic effects and the crucial minimization of $1/f$ noise.

A pivotal technical decision for \textit{LiteBIRD} was the decision to employ a continuously-rotating half-wave plate (HWP) as the first optical element of each telescope in the polarization modulation unit (PMU). The rotating HWP allows us to differentiate between the spurious polarization signal generated within the instrument and the sky polarized signal. In this scheme, the sky signal is exclusively modulated at a frequency equal to $4\times$ the rotation frequency of the HWP ($f_\mathrm{HWP}$). 

\begin{figure}[tp]
    \centering
    \includegraphics[width=.75\textwidth]{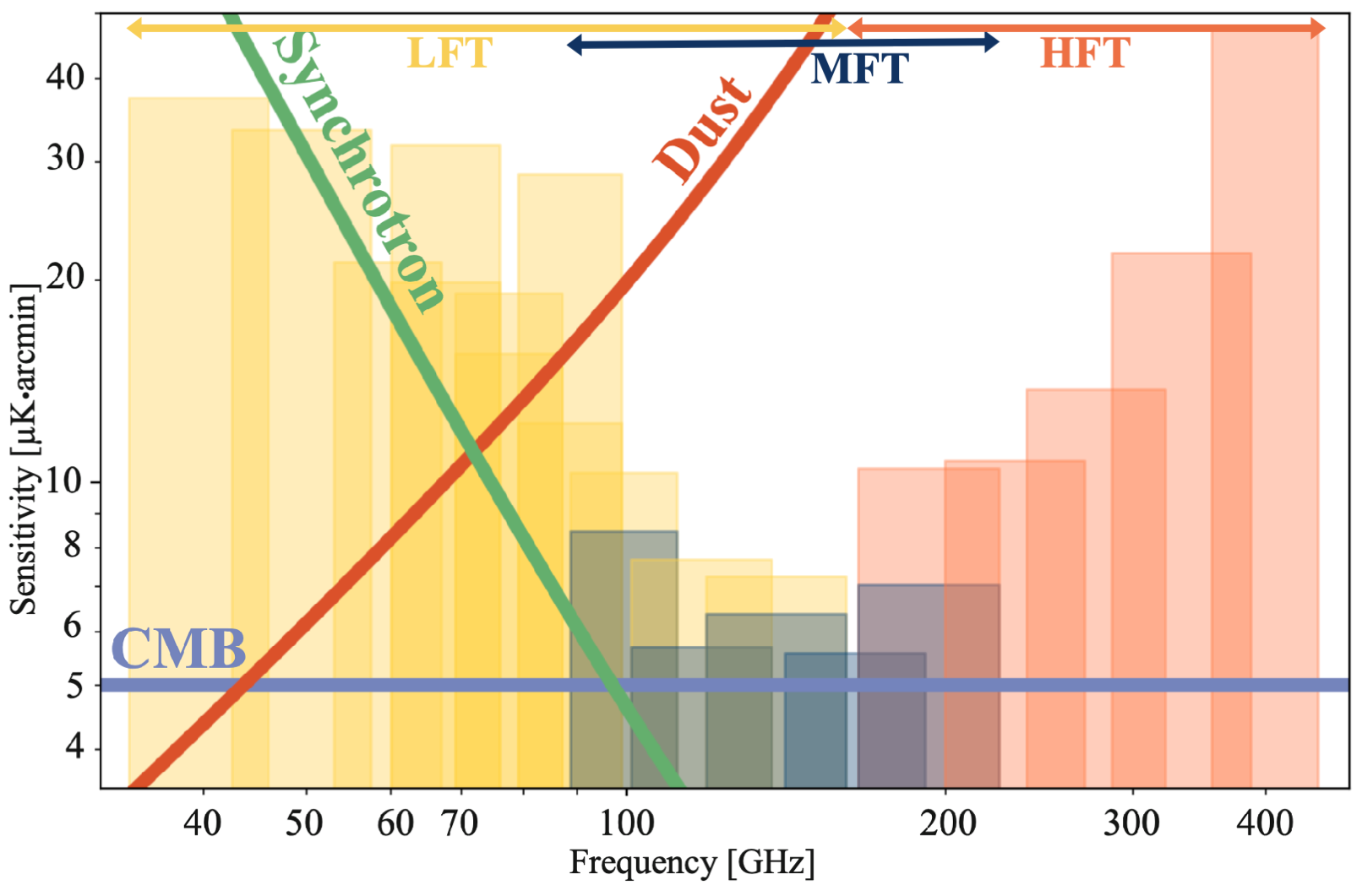}
    \caption{\textit{LiteBIRD} polarization sensitivity per frequency channel color-coded per telescope. For reference we show the expected frequency dependence of CMB, synchrotron, and dust spectra. Figure adapted from \citenum{Hazumi2021PTEP}.}%
    \label{fig:sensitivity}
\end{figure}

This technique can significantly reduce intensity-to-polarization leakage \cite{hoang_2017, Ghigna2020, Patanchon2023}, enabling polarization measurements using a single detector. In addition, the continuously-rotating HWP effectively suppresses $1/f$ noise because of the up-conversion of the signal of interests to a frequency range higher than the $\rm{f_\mathrm{knee}}$ of the $1/f$ noise spectrum. The decision to adopt rotating HWPs is the result of trade-off studies between cases with and without the HWP, including simulations of polarization effects due to HWP imperfections\cite{Hazumi2021PTEP}.

The distribution of bands and their optimization across the \textit{LiteBIRD} frequency range of 34--448\,GHz has been carefully evaluated to address various constraints. The most pressing of these constraints is ensuring appropriate spectral bandwidth and sensitivity to correctly characterize the spectral energy distribution of Galactic foreground sources, namely synchrotron and thermal dust\cite{Fuskeland2023}. However several more issues have been taken into consideration, like optimizing spectral mapping of CO lines\cite{Dame2001, Planck2018I}, and controlling systematic effects through band and instrument overlap that result in redundancy in the data. These trade-off studies resulted in the current baseline configuration with a reflective telescope for low frequencies (LFT, 34--161\,GHz) resulting in a $18^{\circ}\times8^{\circ}$ field-of-view (FoV) and two refractive telescopes for middle and high frequencies (MFT, 89--225\,GHz; HFT, 166--448\,GHz) both with a $28^{\circ}$ (diameter) field-of-view. As visible in Figure~\ref{fig:spacecraft} the telescope mounting configuration in the PLM guarantees coverage of the same circle on the sky with a $180^{\circ}$ lag between LFT and MHFT.

The polarization sensitivity\cite{Hasebe2023} per frequency channel has not changed since the publication of \citenum{Hazumi2021PTEP}, therefore we avoid reporting a complete table here; instead a summary of the sensitivity per frequency channel (color-coded by telescope) can be found in Figure~\ref{fig:sensitivity}.

\begin{figure}[tp]
    \centering
    \includegraphics[width=1\textwidth]{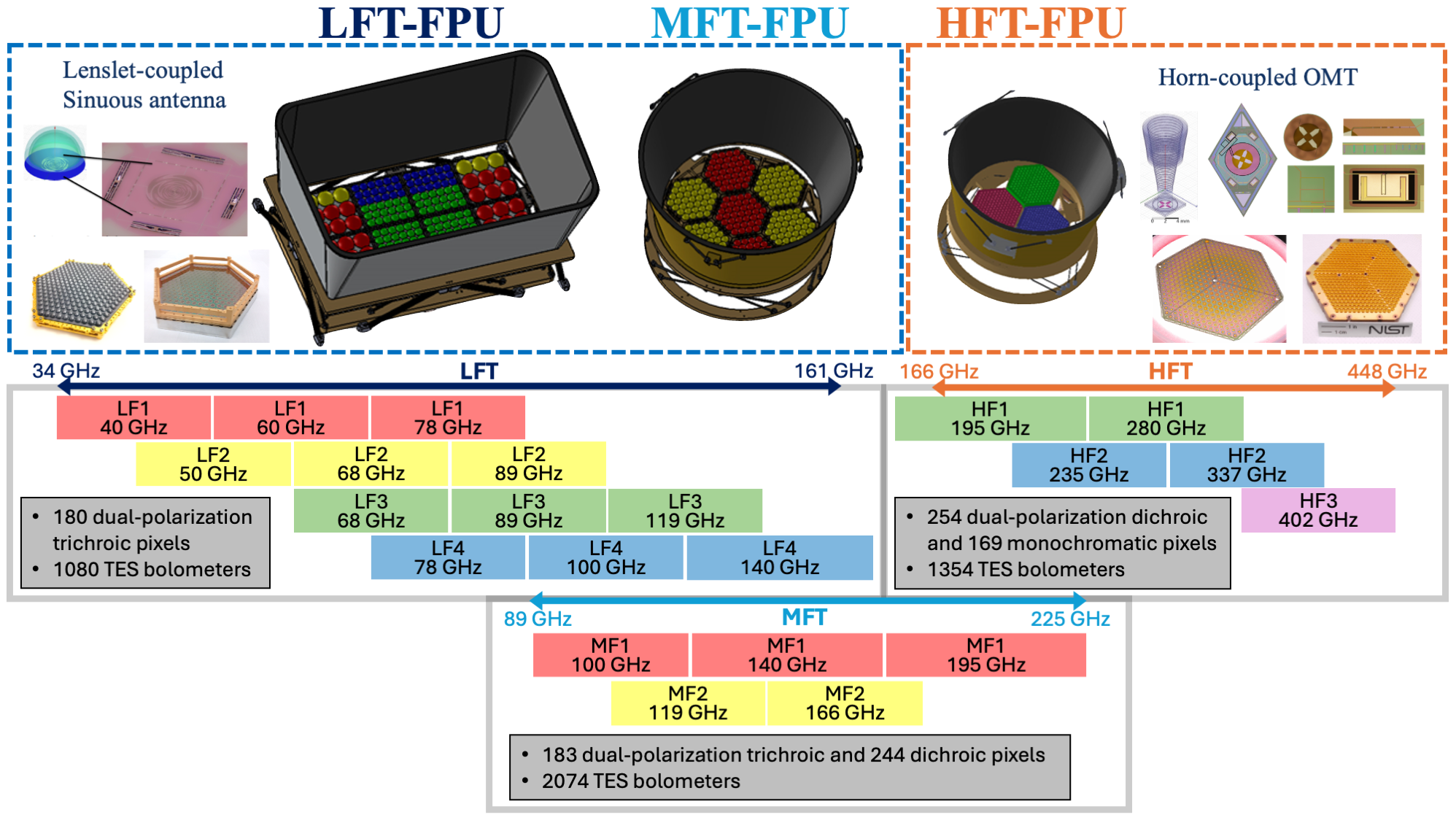}
    \caption{Visual summary of the three focal plane units of \textit{LiteBIRD}. LFT and MFT employ lenslet-coupled sinuous-antenna pixels, while HFT employs horn-coupled OMT (orthomode-transducer) pixels. In the bottom panels we give a visual description of the frequency coverage for the three focal planes and of the frequency channel distribution, as well as the total number and type of pixel and total number of bolometers.}%
    \label{fig:fpu}
\end{figure}

An overiew of the three telescope focal plane units is shown in Figure~\ref{fig:fpu}. The schematic shows the three focal plane units with their respective mechanical assembly, and the distinction between the technology adopted for LFT and MFT, lenslet-coupled sinuous antennas, and the one adopted for HFT, horn-coupled OMTs (orthomode-transducer).
All detectors are polarization sensitive and, with the exception of the highest frequency channel of HFT, they are all multichroic (either dichroic or trichroic) and employ transition-edge sensor (TES) detectors to detect the signal, for a total of 4508 TES bolometers. More details on the design and fabrication of these devices are covered extensively in \citenum{Westbrook2020SPIE, Westbrook2022SPIE, Jaehnig2022SPIE, Hubmayr2022}. Given the large number of detectors to be read together, we have adopted frequency-domain multiplexing as the baseline readout technology.
In particular, the \textit{LiteBIRD} team is making use of the extensive knowledge accumulated by members of our team in the early development of the digital frequency-domain multiplexing (DfMux)\cite{Montgomery2022JATIS} technology.

The \textit{LiteBIRD} project has meticulously planned assembly, integration, verification (AIV), and pre-launch calibration, with a focus on a common calibration approach for the LFT and MHFT instruments. Systematic parameter measurements are derived from detailed systematic effect studies, characterizing performance at the component level and integrating data into an instrument model for in-flight performance forecasting. Instrument-level calibration will be conducted independently in a cold flight-like environment, with ground-calibration operations targeted at ensuring high accuracy for certain parameters. Post-flight analysis mitigation strategies are considered as a potential safeguard, alongside ongoing hardware developments\cite{deHaan2023} and calibration plan refinements throughout the project evolution. A simplified AIV plan is shown in Figure~\ref{fig:req_aiv}c.

\section{OPERATION}
\label{sec:operation}
As already mentioned in Section~\ref{sec:intro}, we will utilize JAXA H3 rocket to launch \textit{LiteBIRD} and put the spacecraft into orbit around the second Sun-Earth Lagrangian point (L2). This location is pivotal to attain a full-sky survey because of its particular location away from the Sun. A Lissajous orbit has been chosen because of its superior thermal conditions compared to a halo orbit.

\begin{figure}[tp]
    \centering
    \includegraphics[width=1\textwidth]{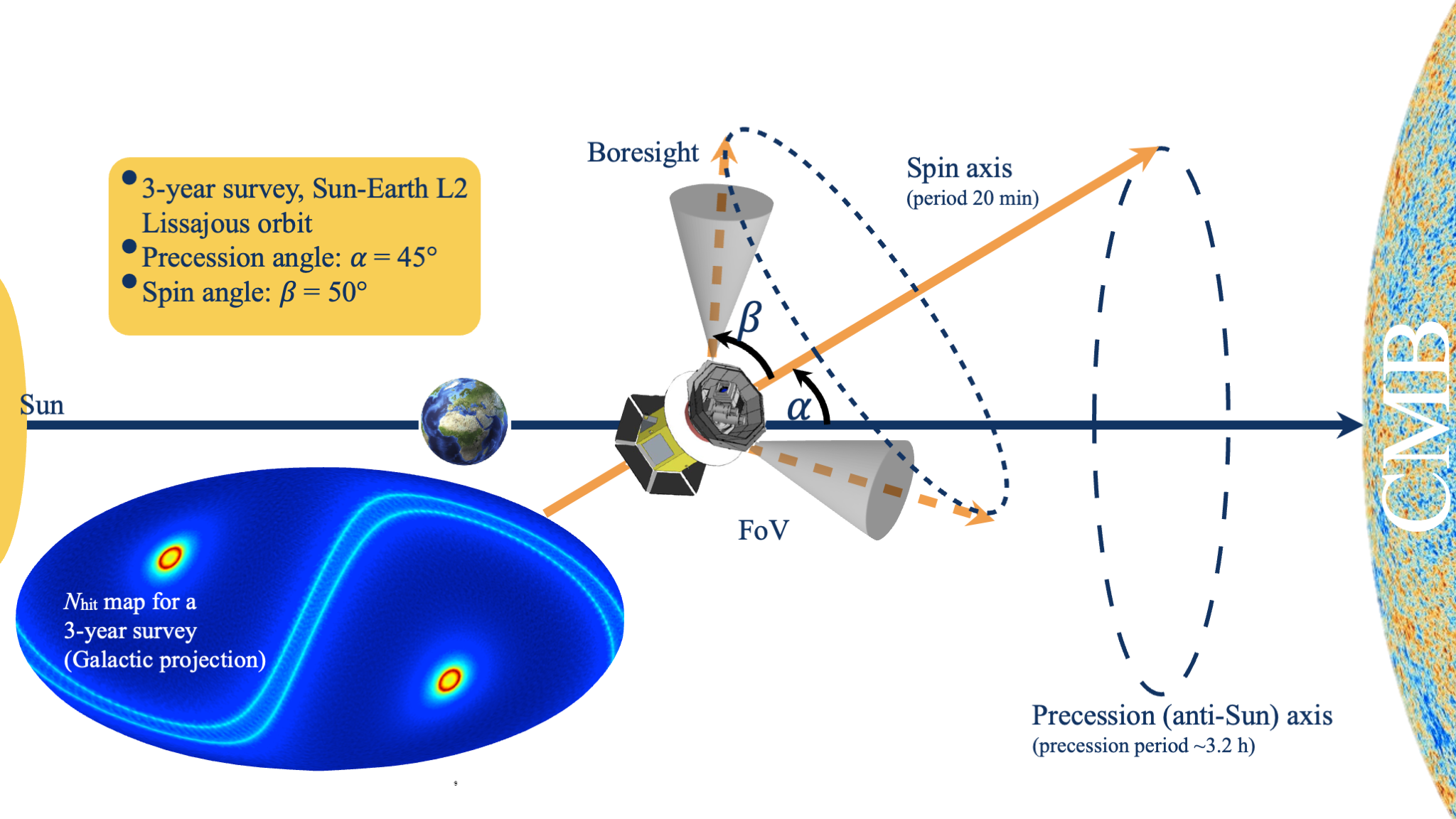}
    \caption{Visual summary of \textit{LiteBIRD} scan strategy, which results from the combination of spin around the major axis of the spacecraft and precession of the spin axis around the anti-Sun direction. In the bottom left the expected hit-map for the whole duration of the mission (3~years) resulting from the baseline scan strategy (combined for all detectors). We can observe the high level of uniformity across most of the sky. Details of the size of the field-of-view (FoV) for all three telescopes can be found in Section \ref{sec:spacecraft}.}%
    \label{fig:scan}
\end{figure}

With the chosen scan strategy (see Figure~\ref{fig:scan}), the spacecraft spins around its major axis at a rate of 0.05\,rpm, equivalent to a period of 20\,minutes. The spacecraft spin is combined with a precession of the spin axis around the anti-Sun direction. The precession rate is much slower than the spin rate and the baseline period has been chosen to be 3.2058\,hours.
This irrational number has been identified and adopted to prevent synchronization issues between the spin and precession that could result in artefacts in the hit-map and cross-linking maps which may give rise to systematic features\cite{hoang_2017}.
The spin axis is inclined at an angle $\alpha = 45^{\circ}$ from the anti-Sun direction, while the angle $\beta$ between the telescopes boresight and the spin axis is $50^{\circ}$.

This scan strategy must meet several requirements, including high thermal stability, uniformity in the distribution of boresight pointing across each sky pixel (cross-linking), observational uniformity across sky pixels (hit-map), large fraction of sky observed daily, and short revisit times for each sky pixel. These requirements are important for mitigation of several instrumental systematic uncertainties.

\section{SCIENTIFIC OUTCOMES}
\label{sec:science}
The \textit{LiteBIRD} collaboration has published in recent years several studies highlighting \textit{LiteBIRD} significant discovery potential, not only when it comes to inflation, but also for other cosmological observables. Details of forecasts of \textit{LiteBIRD} scientific output are presented in \citenum{Hazumi2021PTEP}. However, in the past year more studies have been performed on the lensing analysis\cite{Namikawa2023, Lanoppan2023}, primordial magnetic fields\cite{Paoletti2024}, inflationary models\cite{Campeti2023}, and more.

In doing so \textit{LiteBIRD} members are contributing also to the advancement of data analysis techniques. For example when it comes to foreground cleaning (a pivotal step for cosmological forecasting as well as future data analysis), \textit{LiteBIRD} is the driving force behind several developments to advance both parametric\cite{Poletti2023} and blind component-separation methods\cite{Leloup2023a}. In particular, a lot of emphasis has been put in recent years on techniques that allow us to better account for the complexity of foreground by dividing the sky into patches with the same spectral properties, the so called multi-clustering methods\cite{Puglisi2022, Carones2023, Carones2024}.
Similarly, methods that can take into account the complexity of certain systematic effects are being developed based on the accuracy that will be required by \textit{LiteBIRD}\cite{Leloup2023b, Monelli2023, Patanchon2023}.

For the official cosmological forecasts, we used the polarization sensitivities in Figure~\ref{fig:sensitivity} and applied the method described in \citenum{Errard2019} for foreground cleaning\cite{Hazumi2021PTEP}. Our simulations showed that this design can achieve a total statistical uncertainty $\sigma_\mathrm{stat} = 0.6\times 10^{-3}$ for a tensor-to-scalar ratio $r = 0$ as input.
\begin{figure}[tp]
    \centering
    \includegraphics[width=.75\textwidth]{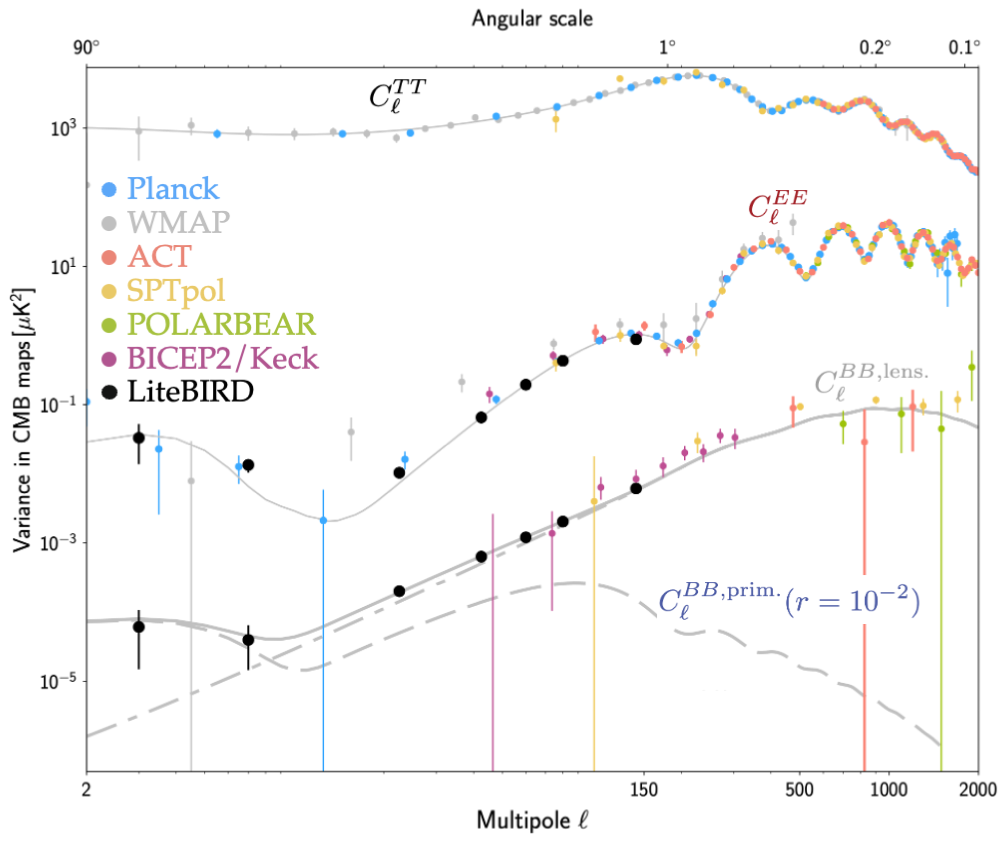}
    \caption{Angular power spectrum summary of the current status of CMB anisotropy results, along with the forecast for \textit{LiteBIRD}. Figure adapted from \citenum{Hazumi2021PTEP}.}%
    \label{fig:spectra}
\end{figure}

As reported in \citenum{Hazumi2021PTEP}, we adopted a systematic approach to estimate systematic uncertainties, identifying 70 items across 14 categories. The \textit{LiteBIRD} team members are constantly working to update this list as new items are identified. We begun our analysis with a simple assumption, where each item is allocated 1\% of the total budget based for systematic uncertainty ($\sigma_\mathrm{syst} = 0.57\times 10^{-3}$).

At present, in our cosmological forecasts, we do not assume combinations of \textit{LiteBIRD} data with external data sets. However, we expect that by combining ground-based projects and space observations better constraints will be obtained. Combining these data, we aim to improve the constraint on the tensor-to-scalar ratio beyond $\delta r < 0.001$ for $2 \leq\ell\leq 200$. For this reason \textit{LiteBIRD} and CMB-S4 have established an MoU (Memorandum of Understanding) to produce joint studies and scientific forecasts.

\textit{LiteBIRD} requirements are derived from tensor-to-scalar ratio measurements alone. Once observations are successfully carried out according to our Level~1 requirements, \textit{LiteBIRD} data will certainly be used to study additional topics in cosmology, particle physics, and astronomy, including for example cosmic birefringence, thermal Sunyaev-Zeldovich effects, elucidating large-angle anomalies, and Galactic astrophysics. 

\section{CONCLUSIONS}
\label{sec:conclusions}
\textit{LiteBIRD} was selected by JAXA in May 2019 as a strategic large-class (L-class) mission. After recently passing one of the many review steps (MDR), it will soon enter the formal pre-project phase (JAXA Phase~A). This phase will last around 3 years and its main focus will be further consolidation of the mission requirements and technological readiness demonstration. At the end of this period the official project phase will begin and the launch is currently expected at the end of Japanese Fiscal Year 2032 (early 2033 in calendar years). The mission is being developed by the \textit{LiteBIRD} Joint Study Group, comprising more than 250 researchers from Japan, North America, and Europe.
\begin{table}[]
    \centering
    \begin{tabular}{|l|l|}
    \hline
    \textbf{Item} & \textbf{Specification} \\
    \hline
    Science requirement & $\delta r < 0.001$ for $2\lesssim \ell \lesssim 200$ \\
    \hline
    Target launch & JFY 2032 (First quarter of 2033) \\
    \hline
    Launch vehicle & JAXA H3 Rocket \\
    \hline
    Observation time & 3\,years \\
    \hline
    Orbit & Sun-Earth Lagrangian point L2 Lissajous orbit \\
    \hline
    Satellite spin & Angle $\beta=50^{\circ}$ -- Period = 20\,minutes \\
    \hline
    Satellite precession & Angle $\alpha=45^{\circ}$ -- Period = 3.2058\,hours \\
    \hline
    PMU revolution rate & 46/39/61\,rpm for LFT/MFT/HFT \\
    \hline
    Sampling rate & 19\,Hz \\
    \hline
    Observing frequency & 34--448\,GHz \\
    \hline
    Number of distinct frequency bands & 15 \\
    \hline
    Polarization sensitivity & 2.2\,$\mu$K-arcmin \\
    \hline
    Angular resolution & $0.5^{\circ}$ (FWHM of LFT 100 GHz) \\
    \hline
    Detectors & Antenna-coupled TES bolometers \\
    \hline
    Readout & Frequency-domain multiplexing (DfMux) \\
    \hline 
    Optics & Reflective cross-Dragone (LFT) -- Refractive (MFT/HFT) \\
    \hline
    Modulation & PMU with sapphire HWP (LFT) -- metalmesh HWP (MFT/HFT) \\
    \hline
    Focal plane temperature & 100 mK (V-grooves -- ST-JT -- ADR cooling chain) \\
    \hline
    Data rate & 17.9\,GB day$^{-1}$ \\
    \hline
    Mass & 3200\,kg \\
    \hline
    Power & 3.7\,kW \\ 
    \hline
    \end{tabular}
    \caption{Summary table of the updated \textit{LiteBIRD} design parameters.}
    \label{tab:paramters}
\end{table}

\textit{LiteBIRD} aims to map the cosmic microwave background (CMB) polarization across the entire sky with unprecedented precision. Its primary scientific goal is to conduct a high-sensitivity search for the signal from cosmic inflation, either confirming its existence or ruling out well-motivated inflationary models. The measurements from \textit{LiteBIRD} have the potential to probe other physics beyond the standard models of particle physics and cosmology.

To meet its essential science requirement of $\delta r < 0.001$, \textit{LiteBIRD} will conduct full-sky surveys for three years from the Sun-Earth Lagrangian point L2. It will utilize 15 frequency bands ranging from 34 to 448\,GHz, divided between three telescopes, achieving a total sensitivity of 2.2 $\mu$K-arcmin and a typical angular resolution of $0.5^{\circ}$ at 100\,GHz.

The project has recently successfully undergone JAXA mission definition review (MDR) and we expect soon an announcement of the official start of Phase~A from JAXA.
Some of the mission parameters have been updated and a summary is given in Table~\ref{tab:paramters}

\acknowledgments   
 
%
This work is supported in Japan by ISAS/JAXA for Pre-Phase A2 studies, by the acceleration program of JAXA research and development directorate, by the World Premier International Research Center Initiative (WPI) of MEXT, by the JSPS Core-to-Core Program of A. Advanced Research Networks, and by JSPS KAKENHI Grant Numbers JP15H05891, JP17H01115, and JP17H01125.
The Canadian contribution is supported by the Canadian Space Agency.
The French \textit{LiteBIRD} phase A contribution is supported by the Centre National d’Etudes Spatiale (CNES), by the Centre National de la Recherche Scientifique (CNRS), and by the Commissariat à l’Energie Atomique (CEA).
The German participation in \textit{LiteBIRD} is supported in part by the Excellence Cluster ORIGINS, which is funded by the Deutsche Forschungsgemeinschaft (DFG, German Research Foundation) under Germany’s Excellence Strategy (Grant No. EXC-2094 - 390783311).
The Italian \textit{LiteBIRD} phase A contribution is supported by the Italian Space Agency (ASI Grants No. 2020-9-HH.0 and 2016-24-H.1-2018), the National Institute for Nuclear Physics (INFN) and the National Institute for Astrophysics (INAF).
Norwegian participation in \textit{LiteBIRD} is supported by the Research Council of Norway (Grant No. 263011) and has received funding from the European Research Council (ERC) under the Horizon 2020 Research and Innovation Programme (Grant agreement No. 772253 and 819478).
The Spanish \textit{LiteBIRD} phase A contribution is supported by MCIN/AEI/10.13039/501100011033, project refs. PID2019-110610RB-C21, PID2020-120514GB-I00, PID2022-139223OB-C21 (funded also by European Union NextGenerationEU/PRTR), and by MCIN/CDTI ICTP20210008 (funded also by EU FEDER funds).
Funds that support contributions from Sweden come from the Swedish National Space Agency (SNSA/Rymdstyrelsen) and the Swedish Research Council (Reg. no. 2019-03959).
The UK  \textit{LiteBIRD} contribution is supported by the UK Space Agency under grant reference ST/Y006003/1 - "\textit{LiteBIRD} UK: A major UK contribution to the \textit{LiteBIRD} mission - Phase1 (March 25)"
The US contribution is supported by NASA grant no. 80NSSC18K0132.
TG acknowledges the support of JSPS KAKENHI Grant Number 22K14054.
%

\bibliography{main} 
\bibliographystyle{spiebib} 

\end{document}